\documentstyle[aps,prl,preprint]{revtex}
\begin{document}
\title {
Quasiparticle spectrum in a nearly antiferromagnetic Fermi liquid:
shadow and flat  bands}
\author{Andrey V. Chubukov}
\address{
Department of Physics, University of Wisconsin, Madison, WI 53706\\
and P.L. Kapitza Institute for
Physical Problems, Moscow, Russia}
\date{today}
\maketitle

\begin{abstract}
We consider a two-dimensional Fermi liquid
in the vicinity of a spin-density-wave
transition to a phase with commensurate antiferromagnetic long-range order.
We assume that near the transition, the Fermi surface is large and
crosses the magnetic Brillouin zone boundary.
 We show that under these conditions, the
self-energy corrections to the dynamical spin susceptibility, $\chi (q,
\omega)$, and to
the quasiparticle spectral function function, $A(k, \omega)$, are
divergent near the transition. We identify and sum
 the series of most singular diagrams, and obtain a solution for
$\chi(q, \omega)$ and an approximate solution for $A(k, \omega)$.
We show that (i) $A(k)$ at a given, small $\omega$ has
an extra peak at  $k = k_F + \pi$
(`shadow band'),  and (ii)
the dispersion near the crossing points is much flatter
than for free electrons.
 The relevance of these results to recent
photoemission experiments in $YBCO$ and $Bi2212$ systems is discussed.
\end{abstract}

\pacs{PACS:  75.10Jm, 75.40Gb, 76.20+q}

The problem of fermions interacting with low-energy magnetic fluctuations
has attracted a considerable interest over the past few years particularly
in connection with high-$T_c$
superconductivity~\cite{hertz,mp,mil,ks,lev2,css,im,at,subir}.
 In this paper, we consider
a two-dimensional system of interacting fermions
 near the antiferromagnetic instability with $Q =(\pi,\pi)$.
 We assume that the Fermi-liquid theory is valid on the disordered side of the
transition (i.e., near the FErmi surface,
$G(k, \omega_m) = Z/(i \omega_m - {\bar\epsilon}_k)$, where ${\bar\epsilon}_k =
\epsilon_k - \mu$, and $Z$ is a positive
constant), and that the Fermi surface is large and crosses the Brillouin zone
boundary - under these conditions,
the transition has a mean-field dynamical exponent
$z=2$~\cite{hertz,mp,mil,css}.
We will show in this paper that due to strong interaction between fermions
and paramagnons,
 the actual form of $G(k, \omega)$ and of the dynamical spin susceptibility
at intermediate energies
is qualitatively different from the prediction of a Fermi-liquid theory.
At the critical point, this new behavior stretches up to
$\omega =0$.

The point of departure for our analysis is
 the spin-density wave (SDW) theory of
an antiferromagnetic transition in a Fermi liquid~\cite{sdw}.
 In this theory, the
instability towards antiferromagnetism occurs when the total magnetic
susceptibility  $\chi (q, \omega) = \chi^0 (q, \omega)/(1 - U_{eff}(q)
\chi^0
(q, \omega))$ diverges at $q=Q, \omega =0$. Here
$\chi^0 (q, \omega)$  is a Pauli susceptibility of an ideal gas (a
particle-hole bubble), and
$U_{eff} (q)$ is an effective interaction. The precise form of
$U_{eff} (q)$ is
irrelevant for our low-energy analysis, we only assume
that $U_{eff} (Q) >0$, and $U_{eff} (Q) Z \leq t$.
A model computation of
$\chi$ in Ref.~\cite{css} yields  $\chi (q, \omega_m) =
  C /(\omega_m^2 + 2 \gamma |\omega_m| +
E^2_{\tilde q})$, where $C$ is a constant of the order of the hopping integral,
$E^2_{\tilde q} = v^2_s {\tilde q}^2 + \Delta^2$, ${\tilde q} = Q - q$.
 This form of
the susceptibility was also suggested on phenomenological
grounds~\cite{david,klr}.
It is essential for our consideraton that the
$\omega_m^2$ and $E^2_{\tilde q}$ terms in $\chi$
come from the integration over the
fermionic momentum in the particle-hole bubble
over the regions far from the Fermi surface. At such scales,
the perturbation theory is non-singular, and we expect that the full
$Re ~\chi^{-1} (q,\omega)$ will not differ qualitatively from the RPA result.
 At the same time, due to energy constraint,
the damping term at $q \approx Q$ comes from the momentum integration over
near vicinity of the crossing points between the Fermi surface and the magnetic
Brillouin zone boundary. At such scales, fermionic and bosonic energies are
both small, and we will show that the perturbation corrections are singular.

Let us first obtain the explicit expression for $\gamma$ in the PRA formalism.
Near each of the crossing  points, the fermionic energies
${\bar \epsilon_k}$ and ${\bar\epsilon}_{k+Q}$
can be expanded quite generally as
 ${\bar \epsilon}_{k} = v{\bar k}
\cos \phi, ~{\bar \epsilon}_{k + Q} =
v{\bar k} \cos (\phi + \phi_0)$, where ${\bar k} = k - k_0$
is the deviation
from the crossing point, $v = (v^2_x + v^2_y)^{1/2}$,
and  for
$v_x >v_y >0$, $\phi_0$ is given by
 $\phi_0 = \pi/2 + 2~\tan^{-1} v_y/v_x$.
A direct calculation of the
particle-hole bubble then yields
$\gamma = U^2(Q)Z^2 C/(\pi v^2 |\sin \phi_0|)$. Notice that $\gamma$
 diverges when $\phi_0 =\pi$ (the $2k_F = Q$ case~\cite{lev}).

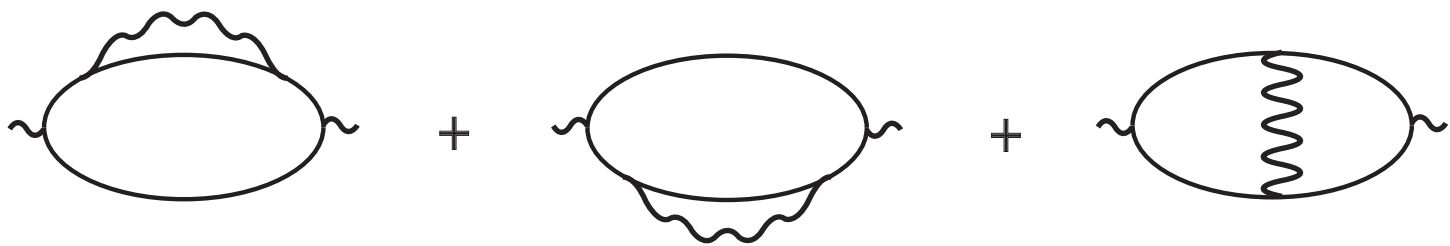
\begin{figure}
\caption{{\em (a)}. Lowest-order self-energy correction due to interaction with
magnetic fluctuations. The solid and wavy lines are fermion and paramagnon
propagators, respectively; {\em (b)}. the
lowest-order corrections to the polarization bubble}
\label{bubble}
\end{figure}

We now go beyond the RPA approximation. Consider first
 whether the $2\gamma |\omega_m|$ term in the magnetic susceptibility
survives the effects of self-energy and vertex corrections at the transition
point.
The lowest-order corrections to the fermionic bubble
are shown in Fig.~\ref{bubble}b.
We explicitly computed the corrections to $\gamma$ from all
three diagrams and
found that
the two diagrams with the self-energy corrections
are free from singularities. However, the diagram with
the vertex correction is logarithmically singular for $Q = (\pi,\pi)$, and
changes $\gamma$ to
${\tilde \gamma} = \gamma (1 - 2 \beta \log (\omega_0/\omega))$ where
$\omega_0$ is the cutoff frequency, and
\begin{equation}
\beta =  \frac{U_{eff}^2 Z^2 C}{4\pi^3 v^2 \gamma}
{}~Re \int_0^{\pi}~d\phi~
\frac{\log[\sin (\phi/2)]}{\cos \phi + \cos \phi_0}
\label{beta}
\end{equation}
Substituting the result for $\gamma$, we find that
$\beta$ in fact depends only on $\phi_0$.
When $\phi_0$ varies between $\pi/2$ and $\pi$,
$\beta$ continuously
varies between $-1/16$ and $-1/8$.
 For $Q \neq (\pi,\pi)$, the logarithmical term is also cut by
$(\pi,\pi) -Q$~\cite{mil}.

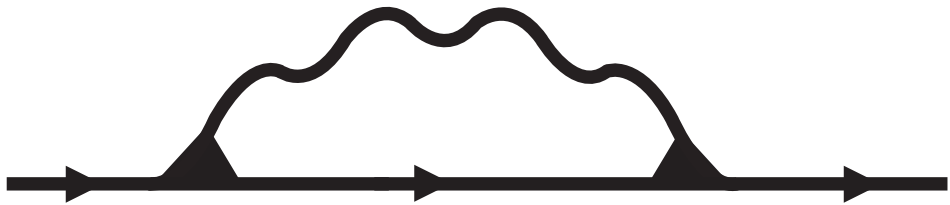
\begin{figure}
\caption {{\em (a)}. Diagram for the
dominant second-order correction to the polarization bubble;
{\em (b)}. diagrammatic expressions for the
fully renormalized polarization bubble and for the full vertex, thick wavy
line is the fully renormalized paramagnon propagator; {\em (c)}. diagram for
the full self-energy. Thick solid line is the full Green function}
\label{bubble2}
\end{figure}

We further considered second order corrections to the polarization bubble,
and found that (i) the dominant contribution comes from the ladder-type
diagram in Fig. \ref{bubble2}a which contributes  $O(\beta^2 \log^2 \omega)$,
while all other second-order diagrams give either finite, or $O(\beta^2
\log \omega)$ contributions, and (ii) the logarithmical terms are cut by the
{\it largest} of the external frequencies. In this situation, the logarithms
sum up to a power law,  and solving the renormalization group (RG)
equation for the
full vertex,  graphically shown in
Fig~\ref{bubble2}b, we obtain $\Gamma \sim \omega^{\beta}$. Substituting this
result into the full bubble (which in our case of the
RG-like perturbation theory contains two full vertices), we
find that
at small frequencies, $\chi^{-1} (q, \omega_m) \propto
{\bar \gamma}~|\omega_m|^{1 + 2 \beta} + E^2_q$, where ${\bar \gamma} \sim
\gamma/\omega_0^{2\beta}$.
An extra logarithmical factor in the frequency term is also possible
due to subleading, double logarithmical corrections to the full vertex
which we didn't compute. We see that the functional form of the
full dynamical susceptibility at the transition point
is different from the RPA and phenomenological predictions, though the
numerical value of $\beta$ turns out to be small~\cite{beta}. Away from the
transition, the logarithmical singularities are cut by $\Delta$, and
there is a crossover, at
$|\omega_m| \sim \Delta^2/\gamma$,  from
an anomalous $|\omega_m|^{1 + 2 \beta}$ dependence of $\chi$
at higher frequencies to
a Fermi-liquid-type frequency dependence $|\omega_m| \Delta^{4 \beta}$ at
the lowest frequencies.

Having obtained the result for the spin susceptibility, we
 now turn to the discussion of the form of the full
fermionic Green function.
Consider first the  lowest-order self-energy
correction to the fermionic propagator (Fig.~\ref{bubble}a). We have
 $G^{-1} = (i \omega_m -
\bar{\epsilon}_k - \Sigma (k, \omega_m))/Z$, where
$\Sigma (k,  \omega_m) = Z U^2(Q)
\int \chi (q, \Omega_m) G ({k} + q, \Omega_m + \omega_m)$,
 where $\chi (q, \Omega_m)$
is the RPA susceptibility.
The evaluation of the
self-energy is tedious but straightforward. We obtained
\begin{eqnarray}
&&\Sigma (k, \omega_m) =
 - i \omega_m~
\left(\frac{2 \gamma}{|\omega_m|}\right)^{1/2} \frac{v~|\sin
\phi_0|}{4v_s}~\times \nonumber \\
&&\Phi\left(\frac{\Delta^2}{2 \gamma |\omega_m|},
 \frac{v^2~{\bar \epsilon}^{2}_{k+Q}}{2 v^2_s ~\gamma |\omega_m|}\right) ~-~
\frac{L |\sin \phi_0|}{4 \pi}~\left({\bar \epsilon}_{k + Q} - i
\omega_m\right)
\label{s-e}
\end{eqnarray}
where $ L = \log [\omega_0/max (\omega, \Delta, \bar \epsilon)]$.
The function
$\Phi(x,y)$ has a simple form in the two limits:
 at the crossing point, ${\bar \epsilon}_{k + Q} =0$,
we have $\Phi (x,0) = (x^{1/2} + (x + 1)^{1/2})^{-1}$, and at the transition
point, $\Delta =0$, we found that it is
 well approximated by $\Phi (0,y) =
 \pi^{-1} (1 + y)^{-1/2} \cot^{-1} ((y-1) a/(y+1)^{3/2})$, where
$a \sim (2 \gamma/|\omega_m|)^{1/2} \gg 1$.

The key observation from eq. (\ref{s-e}) is that the self-energy correction
to the quasiparticle Green function at ${\bar k} = 0$
has a power-law singularity: $\Sigma (k_0, \omega_m) \propto
i \omega_m/|\omega_m|^{1/2}$, which
for $\omega \ll \gamma$
clearly overshadows the zero-order, $i \omega_m$ term in $G^{-1}_0$.
At the same time,  the self-energy correction
at zero frequency is only logarithmically divergent~\cite{comm}. The
singular behavior of $\Sigma (\omega)$
was first obtained analytically by Millis~\cite{mil}, and then
used in the `spin gap' calculations in~\cite{lev2}. Numerically,
the analogous frequency dependence was obtained by
 Monthoux and Pines~\cite{mp}.
The logarithmical divergence of the self-energy
at zero frequency has not been studied before, to the best of our
knowledge. Notice also that (\ref{s-e}) does not lead to the shift
in the location of the crossing point.

We now discuss the form of the full $G(k,\omega_m)$.
The diagram for the full self-energy
is shown in Fig~\ref{bubble2}c.
It contains two full vertices,
 full spin propagator, and full
fermionic Green function.
We found above that the lowest-order
corrections to vertices and spin propagator contain logarithms while
the correction to the fermionic propagator contains both square-root and
logarithmical singularities. Because of three different sources for
logarithms, the explicit summation of the perturbation
series is hardly possible. Below we obtain an approximate solution
for $G(k, \omega_m)$: we neglect all logarithmical terms but keep power-law
divergencies. This approximation is not fully self-consistent,
as the series of
logarithms may eventually give rise to extra powers of frequency
and momentum, similar to what we found above for the
spin propagator. However, the
lowest-order logarithmical corrections contain small numerical factors
($\beta \ll 1$ for susceptibility and vertex function,
and $|\sin \phi_0|/4\pi$ for the Green function, see
eq. (\ref{s-e})), so  the logarithmical terms are likely to be irrelevant
for all practical purposes.

We now proceed with the calculations. Assume that the fully
renormalized fermionic propagator has a form
$ G( k, \omega_m) = S({k}, \omega_m)/(i \omega_m
|{\bar\omega}_0 /\omega_m|^{1/\alpha} - {\bar \epsilon}_{k})$.
Substituting this Green function into the self-energy term, we obtain after
simple manipulations   that at the
transition point, $\Sigma (0, \omega_m) \propto S~(i \omega_m)
|{\bar\omega}_0/\omega_m|^{1/2}$ {\it independent} on $\alpha$.
 At the same time, at zero frequency, we find, neglecting logarithms,
$\Sigma (k, 0) \propto S~ {\bar
\epsilon}_{k + \pi}$.
Self-consistency on $G(k, 0)$ then implies
 that $S = O(1)$.
Substituting $S$ into a self-consistency condition on $G(k_0, \omega_m)$,
 we  obtain  $\alpha = 2$.

The extension of the above arguments to
$\Delta \neq 0$ is straightforward, and we finally obtain
\begin{equation}
G(k, \omega_m) = \frac{Z}{i \omega_m
|\frac{{\tilde\omega}_0}{\omega_m}|^{1/2}
{}~{\tilde \Phi}\left(\frac{\Delta^2}{2 \gamma |\omega_m|},
{}~\frac{v^2~{\bar \epsilon}^{2}_{k+Q}}{2 v^2_s \gamma |\omega_m|}\right)
{}~- {\bar \epsilon}_{k}},
\label{gtot}
\end{equation}
where $\tilde\omega_0 \sim \gamma v^2/v^2_s \sim \omega_0 \sim t$,
 and the function $\tilde \Phi$ has the same asymptotic behavior
as $\Phi$ introduced earlier  (in particular, $\tilde \Phi (0,0) =1$),
 but  may differ from  $\Phi$ at intermediate values of arguments.
Eq. (\ref{gtot}) is the key result of this paper.

We now discuss two applications of this result relevant to experiments.
First, we show that eq. (\ref{gtot}) yields
flat quasiparticle dispersion  near each of the crossing points.
In the
photoemission experiments, the quasiparticle energy is associated
with the position of the maximum in the spectral function $A(k ,\omega)$
at a given $k$. In a Fermi-liquid,
$A (k, \omega)$  has a  peak at $\omega = {\bar \epsilon}_k \propto
v|k - k_0|$, where $k_0$ is one of the crossing points.
 The full spectral function near
$k_0$, which one obtains from (\ref{gtot}) after a transformation to real
frequencies, however,
has a maximum  at
$\omega = B_k~{\bar\epsilon}^2_k/\omega_0~
 \propto (k -
k_0)^{2}$ at $\Delta =0$, and at $\omega \propto \Delta (k - k_0) +
O((k-k_0)^2)$
in the disordered phase.
The factor $B_k$ depends on the ratio
${\bar \epsilon}_{k+\pi}/{\bar \epsilon}_{k}$,
 but we have checked numerically
that this dependence is actually very weak.
We see that for $\Delta \ll v \sim \omega_0$,
 which we assume to hold near the magnetic
transition, the effective quasiparticle dispersion near $k_0$
 is nearly quadratic rather than linear, which obviously means
 that it is much flatter than the dispersion of free fermions.

Another application relevant to experiments is the appearance of a
`shadow band' near the transition.
Suppose that we are
some distance away from $k_0$, such that
 self-energy corrections are non-singular.
In this situation, $A(\omega)$ at a given
$k \approx k_F$ is dominated by a conventional quasiparticle peak, i.e.,
$A(\omega) \propto Z {\tilde \gamma} \omega^2 /
({\tilde \gamma}^2 \omega^4 +
(\omega - {\bar \epsilon}_k)^2)$.
Right at the Fermi surface we have $A(0) >0$,
while at $k \neq k_F$, $A(\omega)$
 behaves as $\omega^2$ at the lowest frequencies,
and has a sharp peak at $\omega = {\bar \epsilon}_k$.
 As we move away from the Fermi surface, the peak
shifts to higher frequencies, and the low-frequency part of $A(\omega)$
flatters.
This behavior breaks down however when $k$ reaches the value $k = k_F + Q$.
At this point,  ${\bar \epsilon}_{k + Q} =0$, and the
 self-energy term has the same singularity
at small $\omega$ as in eq. (\ref{s-e}), i.e.,  at $\Delta =0$ we have
 $\Sigma (\omega_m) \propto i\omega_m/|\omega_m|^{1/2}$~\cite{mil,lev2}.
Doing the standard manipulations, we  obtain at  small frequencies,
 $A(\omega >0)
\sim \omega^{1/2}/\omega_0^{3/2}$ rather than $A (\omega) \sim \omega^2
/\omega_0^3$ which holds
when both
${\bar \epsilon}_k$ and  ${\bar \epsilon}_{k + Q}$ are finite. If $\Delta>0$,
 the quadratic dependence exists also at
$k = k_F +Q$, but at these $k$,
$A(\omega) \propto \omega^2/\Delta^{3}$,
i.e., for $\Delta \ll \omega_0$, the slope is still substantially larger than
for other momenta.
 The increase in the slope of $A(\omega)$
at $k = k_F +Q$ can be detected if
 one fixes $\omega$ at some small value, and plots $A$ as a
function of $k$. Clearly, $A(k)$
 should have two maxima: one where ${\bar \epsilon}_k = \omega$, and
the other (smaller and  broader) where ${\bar \epsilon}_{k + Q} =0$. The
second peak is usually referred to as a `shadow band'~\cite{ks}.
 Notice however the important
difference between conventional and `shadow' peaks: at ${\bar \epsilon}_k
=0$, $A(\omega)$ tends to a finite value at $\omega =0$, while at
${\bar \epsilon}_{k + Q} =0$, we  have
$A(0) \equiv 0$. This argument shows that in the absence of long-range
magnetic order, the `shadow Fermi surface', striktly speaking, does not
exist~(cf. Ref\cite{chak}).
 In practice, however, the measurements always involve averaging over some
finite frequency range due to resolution,
in which case the second peak in $A(k)$ should indeed be
present in the data.

We now discuss possible experimental realizations of these effects.
First,  recent photoemission measurements of
 $A(k)$ at {\it small} $~\omega$
in pure and lead-doped $Bi2212$~\cite{aebi}  have shown that
the  intensity has a second, `shadow' peak located at $k = k_F + Q$, where
${\bar \epsilon}_{k+Q} =0$.
This is totally consistent with our findings. Second,
the flat quasiparticle dispersion  near $(0,\pi)$
has been observed in nearly optimally doped $YBCO$~\cite{gofron}
and in $Bi2212$~\cite{ma,shen}.  In both
systems (particularly in $Bi2212$)
the Fermi surface crosses the magnetic
Brillouin zone boundary near $(0,\pi)$ and symmetry related
points~\cite{ma,aebi,liu}. Assume that
${\bar \epsilon}_{(0,\pi)} \ll \omega_0$, i.e, eq.
(\ref{gtot}) is valid near $(0,\pi)$. Then the effective quasiparticle
dispersion
 is $E_k  = E_{(0,\pi)} + \delta ~\Delta \epsilon_k$, where
$\Delta \epsilon_k =
\epsilon_k -\epsilon_{(0,\pi)},~\delta \sim
({\epsilon}^2_{(0,\pi)} + \lambda~\Delta^2)^{1/2} /\omega_0,~
\lambda = O(1)$, and $E_{(0,\pi)} - \mu \propto
{\bar\epsilon}^2_{(0,\pi)}/\omega_0
\ll {\bar\epsilon}_{(0,\pi)}~$.
 We see that if $\epsilon_{0,\pi}$ and $\Delta$ are both substantially smaller
than $\omega_0 \sim t$, then $\delta \ll 1$, and
the actual quasiparticle dispersion near $(0,\pi)$ has
an extra small factor compared to the
mean-field dispersion.  Note, however, that
 in $YBCO$, the measured Fermi surface was fitted to the
$t-t^{\prime}$ model with $t^{\prime} =
-0.45 t$~\cite{si}.
In this model, the dispersion of free fermions is already flat:
$\Delta \epsilon_k = - 0.1 t~(\pi -y)^2$ for $x=0$, and
$ \Delta \epsilon_k = 1.9 t~x^2$ for $y=\pi$.
We see that the dispersion
along $(0,y)$ (but not along $(x,\pi)$) is flat already at the mean-field
level, and  it is actually difficult to judge to
which extent the measured flat band is due to the effects of the interaction,
 and to which
extent it is  a property of a  dispersion of free fermions
near the Fermi surface.
In $Bi2212$, the value of $t^{\prime}$ is unknown, but
the measured location of the crossing point $k_0$
 is closer to $(0,\pi)$ than in $YBCO$~\cite{ma,aebi},
so we expect our theory to be more relevant.
There is a clear indication from the recent data~\cite{ma} that the dispersion
around $(0,\pi)$ is flat along both $(0,y)$ and $(x, y_0)$ directions, where
$y_0$ is close to $\pi$. This
phenomenon is
consistent with the scenario of the fluctuation-induced softening.

To summarize, we found that the interaction between fermions and
low-energy spin fluctuations
strongly affects the frequency-dependent part of the
the fermionic quasiparticle Green function.
To lesser extent, the renormalization also affects the imaginary part of the
staggered spin susceptibility, and the momentum-dependent
part of the fermionic propagator.
 The full spectral
function $A(\omega)$ at a given $k$ located near the point
where the Fermi surface crosses the
magnetic Brillouin zone
boundary, has a peak at $\omega \sim (\epsilon_k -\mu)^{2}$
(up to logarithms),
i.e., fluctuation corrections
flatten the quasiparticle dispersion.
The spectral function $A(k)$ measured as a function of momentum
at a given (small) frequency
has two peaks: a conventional  quasiparticle peak,
and a `shadow' peak located where $\epsilon_{k+Q} =\mu$.

The results above are related to other
works on shadow and flat bands. The `shadow bands' in a nearly
antiferromagnetic Fermi liquid were first
obtained by
Kampf and Schrieffer~\cite{ks}.
 Our approach is similar to theirs except that
we considered here a conventional `small $U$' SDW
transition, when the SDW gap is zero at the transition
point and gradually increases in the ordered phase, while Kampf and
Schrieffer considered semiphenomenologically the `large $U$' limit near
half-filling,
when there exists a pseudogap in
the electronic spectrum in the paramagnetic phase.
Dagotto et al~\cite{dagotto}  argued recently  that
the shadow and flat bands both have antiferromagnetic origin.
 Our conclusions
are consistent with theirs. They however
related the measured flat dispersion near $(0,\pi)$ to the flatness of the
spectrum of a single hole in a half-filled $t-J$ model, which,
they argue, persists at finite doping. We
 have shown here
that there exists an additional, model independent mechanism
which flatters  the quasiparticle dispersion  near
the magnetic transition.\\
{\it Note added}. After this paper was completed, I received a preprint from
B.L. Altshuler, L.B. Ioffe and A.J. Millis~\cite{lev} on the analysis of the
SDW transition at $Q = 2k_F$. They consider a case $Q \neq (\pi,\pi)$, but also
discuss scaling behavior at intermediate momenta when the difference between
$Q$ and $(\pi,\pi)$ can be neglected. In this regime, they identify the
series of logarithmical corrections, and their
 results are very similar to the ones presented here.

It is my pleasure to thank S. Sachdev for numerous discussions and comments.
 I am also thankful to P. Aebi, E. Dagotto, D. Frenkel,
L. Ioffe, R. Joynt, Jian Ma, A. Millis, A. Moreo,
M. Onellion, D. Pines, A. Sokol  and C. Varma for useful conversations.
The author is an A.P. Sloan fellow.

\begin{references}
\bibitem{comm} One may wonder why the singular self-energy correction to the
fermionic propagator does not show up in the renormalization of the $\gamma
\omega$ term in the susceptibility. The argument here
is that the damping term
depends on the ratio $Z/v$. The singular $\sqrt \omega$
 frequency correction to $G$ equally renormalizes $Z$ and $v$,
 and therefore leaves the ratio unchanged.
\end{references}
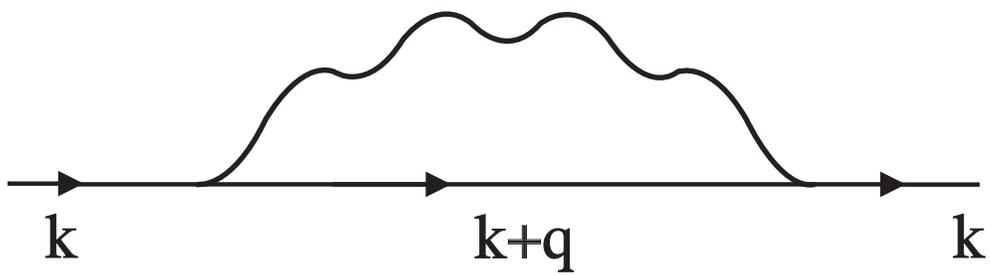
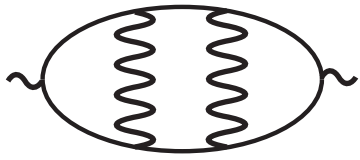
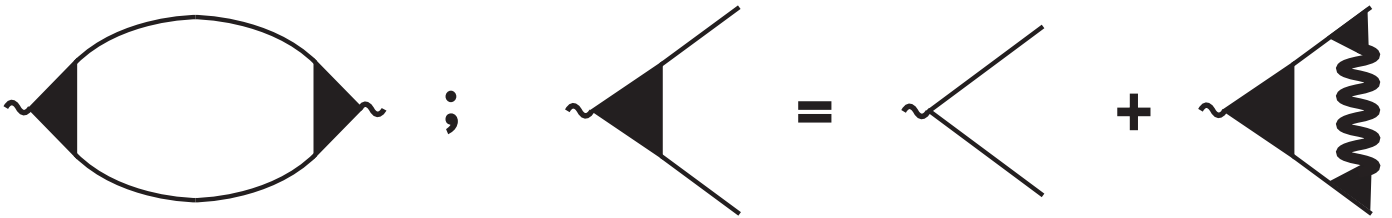
\end{document}